\documentclass[twocolumn,floatfix,amsmath,amssymb]{revtex4}
\usepackage{graphicx}
\usepackage{dcolumn}
\usepackage{color}

\newcommand{\be}{\begin{equation}}
\newcommand{\ee}{\end{equation}}

\begin{document}
\title{Intermittency in Hall-magnetohydrodynamics with a strong guide
field}

\author{P.~Rodriguez Imazio$^1$, L.N.~Martin$^1$, P.~Dmitruk$^1$, and 
        P.D. Mininni$^{1,2}$}
\affiliation{$^1$ Departamento de F\'{\i}sica, Facultad de Ciencias Exactas y 
                  Naturales, Universidad de Buenos Aires and IFIBA, CONICET, 
                  Buenos Aires 1428, Argentina \\
             $^2$ National Center for Atmospheric Research, P.O.~Box 3000, 
                  Boulder, Colorado 80307, USA}
\date{\today}

\begin{abstract}
We present a detailed study of intermittency in the velocity and magnetic 
field fluctuations of compressible Hall-magnetohydrodynamic turbulence 
with an external guide field. To solve the equations numerically, a 
reduced model valid when a strong guide field is present is used. 
Different values for the ion skin depth are considered in the simulations. 
The resulting data is analyzed computing field increments in several 
directions perpendicular to the guide field, and building structure 
functions and probability density functions. In the magnetohydrodynamic 
limit we recover the usual results with the magnetic field being more 
intermittent than the velocity field. In the presence of the Hall effect, 
field fluctuations at scales smaller than the ion skin depth show a 
substantial decrease in the level of intermittency, with close to 
monofractal scaling.
\end{abstract}
\maketitle

\section{Introduction}

The properties of small scales structures in magnetohydrodynamic (MHD) 
and Hall-magnetohydrodynamic (HMHD) turbulence have been the subject of 
conflicting results and of several debates. In particular, much 
attention has been paid in the literature to the geometrical properties 
of current sheets in HMHD, as these structures are associated with 
magnetic flux reconnection and magnetic energy dissipation, processes 
of uttermost importance in astrophysics and space physics 
\cite{Birn01,Shay01,Ren05,Dmitruk06}.

While some numerical simulations indicate that current sheets affected 
by the Hall effect are wider than in MHD (see, e.g., \cite{Martin12}), others 
observe thinner structures \cite{Donato1} In all cases, the geometry of the 
currents sheets is changed, displaying the so-called X-type structure 
and reminiscent of the Sweet-Parker configuration in the MHD case 
\cite{Servidio09}, and changing to a double wedge shape reminiscent 
of the Petschek configuration when the Hall effect is relevant 
\cite{Yamada06}. In simulations of turbulent HMHD, it was observed 
that the peak of the spectrum of the current density was located at a 
wavenumber corresponding to the inverse of the ion skin depth 
\cite{Mininni02,Mininni03,Mininni05,Mininni07}. Since this peak can be 
associated with an average thickness of the current sheets, the effect 
was interpreted as a thickening of the current sheets as the Hall 
effect was increased \cite{Gomez10}. The result is in good agreements 
with experiments that indicate that the thickness of the current sheet 
in the presence of the Hall effect is given by the ion skin depth 
\cite{Yamada06}.

Ref.~\cite{Martin12} provides a possible answer to these conflicting 
results. In simulations of turbulent HMHD with a guide field, the
authors observe that although the current sheet widens as the ion 
skin depth is increased, it also fragments internally into smaller 
filaments. 

The case in which thinner structures were observed \cite{Donato1}
suggests that HMHD is more intermittent than MHD. This is also the 
case in some observations in the solar wind turbulence using the
Cluster magnetic data \cite{Alexandrova07,Alexandrova08}. 
However, other observations in the solar wind of high-frequency 
magnetic field fluctuations from the same spacecraft indicate that 
while large scales are compatible with multifractal intermittent 
turbulence, small scales show non-Gaussian monoscaling 
\cite{Kiyani09}.

A quantification of the level of intermittency is important to 
understand the geometrical distribution of dissipation in a 
magnetofluid and a plasma, and it also can provide constraints 
for theories of magnetic energy dissipation and reconnection. While 
previous analysis of intermittency in HMHD were mostly based on 
the differences observed in the geometry and size of current sheets,
or in the study of probability density functions (PDFs) of field 
increments at different scales, a precise quantification requires 
computation of both PDFs and of structure functions.

The study of intermittency based solely on observations of individual 
structures has several shortcomings. Although the formation of 
small scale structures can point out to an increase in the level of 
intermittency, there is more information that is needed to make such 
claim. If there are thinner structures, are these structures spatially 
localized? Or do they occupy more space than in the MHD case, thus 
being space filling? In the former case, HMHD would be more 
intermittent, while in the latter case intermittency would be
decreased by the Hall effect.

In this work we present a detailed study of intermittency in the 
velocity and magnetic field fluctuations. Considering the solar wind 
as a motivation, the data for the analysis stems from numerical 
simulations of MHD and HMHD turbulence with a guide field. We use the 
reduced MHD (RMHD, \cite{Strauss76,Montgomery82}) and reduced HMHD 
(RHMHD, \cite{Gomez,Martin11}) models to generate data under the 
approximation of a strong guide field. Then, structure functions and PDFs 
of the fields are computed, for increments in the direction perpendicular 
to the guide field. To reduce errors, an average of the structure 
functions for several directions perpendicular to the guide field 
is computed using the SO(2) decomposition \cite{Mininni10,Imazio11}. 
Although at small scales in the solar wind several kinetic effects 
may play important roles, we found that a simple Hall magnetofluid 
reproduces some of the the observations in \cite{Kiyani09}, and that 
the presence of the Hall effect results in a substantial decrease in 
the intermittency of the velocity and magnetic fields at scales 
smaller than the ion skin depth.

\section{Reduced MHD and HMHD models}

For a compressible flow, the HMHD equations can be written (in 
dimensionless form) as
\begin{eqnarray}
\frac{\partial{\bf{u}}}{\partial t} - {\bf{u}} \times \boldsymbol{\omega}= 
  -\nabla \left(\frac{{\bf{u}}^{2}}{2} + 
  \frac{\rho^{\gamma-1}}{M_{S}^2(\gamma-1)} \right) + {}
  \nonumber\\
{} {}             
  + \frac{1}{M_{A}^{2}} \frac{\bf{J} \times \bf{b}}{\rho} 
  +\nu \frac{\nabla^{2}{\bf{u}}}{\rho} + \left(\delta+\frac{1}{3}\nu
  \right) \frac{\nabla ({\nabla \cdot \bf{u}})}{\rho},
\label{eq:1}
\end{eqnarray}
\begin{equation}
\frac{\partial{\bf{A}}}{\partial t} - {\bf{u}}\times \bf{b} = 
  - \epsilon \frac{\bf{J} \times \bf{b}}{\rho} -\nabla \phi +
  \eta \nabla^2\bf{A},
\label{eq:2}
\end{equation}
\begin{equation}
\frac{\partial{\rho}}{\partial t} + \nabla \cdot (\rho {\bf{u}}) = 0,
\label{eq:3}
\end{equation}
\begin{equation}
\nabla \cdot {\bf{A}}=0.
\label{eq:4}
\end{equation}
In these equations, ${\bf u}$ is the velocity field, ${\boldsymbol\omega}$ 
is the vorticity field, ${\bf J}$ is the current, ${\bf b}$ is the 
magnetic field, $\rho$ is the density of the plasma, and ${\bf A}$ and 
$\phi$ are respectively the magnetic and electric potentials. A 
barotropic law is assumed for the plasma, with the pressure given by 
$p=C \rho^{\gamma}$, where $C$ is a constant and $\gamma=5/3$. Equation 
(\ref{eq:4}) is the Coulomb gauge, which acts as a constraint that fixes 
the electric potential in Eq.~(\ref{eq:2}). Control parameters of the 
system are the sonic Mach number $M_S$, the Alfv\'en Mach number $M_A$, 
the viscosities $\nu$ and $\delta$ (here we consider $\nu=\delta$),
and the resistivity $\eta$. In our study, the most important control 
parameter is the Hall coefficient $\epsilon = \rho_{ii}/L$, where
$\rho_{ii}$ is the ion skin depth and $L$ is the characteristic scale
of turbulence. When $\epsilon=0$, the equations above result in the
well known compressible MHD equations.

In the presence of a strong guide field, the equations above can be 
written using the reduced approximation often used in 
magnetohydrodynamics (see, e.g., \cite{Strauss76,Montgomery82}). The 
approximation assumes that the magnetic field can be written as
\begin{equation}
{\bf b} = B_0 {\bf{\hat{z}}} + {\bf b}',
\end{equation}
where $B_0$ is the intensity of the guide field, and ${\bf b}'$ 
is such that $|{\bf b}'|/B_0 \ll 1$.

We assume, without loss of generality, that $B_0 = 1$, and we 
decompose the velocity and magnetic field fluctuations in terms of 
scalar potentials as
\begin{equation}
{\bf u} = \nabla \times \left( \varphi {\bf{\hat{z}}} + 
  f {\bf{\hat{x}}} \right) + \nabla \psi ,
\label{u1}
\end{equation}
and
\begin{equation}
{\bf b}' = \nabla \times \left( a {\bf{\hat{z}}} + g {\bf{\hat{x}}} \right).
\label{b1}
\end{equation}
Equation (\ref{b1}) ensures that the magnetic fields remains
divergence free, while Eq.~(\ref{u1}) gives us a compressible flow. 
The potentials $f$ and $g$ allow for dynamical components of the 
fields parallel to the guide field, and $\psi$ describes an
irrotational component of the velocity field.

Then, Eqs.~(\ref{eq:1}-\ref{eq:4}) can be written as (for the details 
see \cite{Gomez} and \cite{Martin11,Bian,Martin12})
\begin{equation} \label{rhmhd1}
\frac{\partial u}{\partial t}= \frac{\partial b}{\partial z} +
[\varphi,u] - [a,b] + \nu \nabla^2 u,
\end{equation}
\begin{equation} \label{rhmhd2}
\frac{\partial \omega}{\partial t}= \frac{\partial j}{\partial z} + 
[j,a] - [\omega,\varphi] + \nu \nabla^2 \omega,
\end{equation}
\begin{equation} \label{rhmhd3}
\frac{\partial a}{\partial t}=
\frac{\partial (\varphi-\epsilon b)}{\partial z} +
[\varphi,a] - \epsilon[b,a] + \eta \nabla^2 a,
\end{equation}
\begin{eqnarray} \label{rhmhd4}
\frac{\partial b}{\partial t}= \beta_p\frac{\partial (u-\epsilon j)}{\partial z}
+ [\varphi,b] + \beta_p[u,a]+\nonumber\\ - \epsilon \beta_p[j,a] + 
\eta \beta_p \nabla^2b,
\end{eqnarray}
where
\begin{eqnarray}
u=-\partial_y f\label{u},\\
\omega=-\nabla^2_\perp \varphi\label{omega},\\
b=-\partial_y g\label{b},\\
j=-\nabla^2_\perp a\label{j},
\end{eqnarray}
and the notation $[A,B]=\partial_x A \partial_y B - \partial_x B \partial_y A$
is employed for the Poisson bracket. The potential $\psi$ was
eliminated from these equations using the equation for the pressure. 
Finally, $\beta_p=\beta\gamma/(1+\beta\gamma)$ is a function of the 
plasma ``beta''. As in the previous set of equations, these equations 
become the compressible RMHD equations when $\epsilon=0$.

\section{Numerical simulations}

Simulations analyzed in this work are similar to those described in 
Ref.~\cite{Martin12}. We use a standard parallel pseudospectral code to 
evaluate the nonlinear terms and solve numerically the equations 
\cite{Ghosh93}. A second-order Runge-Kutta time integration 
scheme is used. The magnetic field fluctuations in all simulations 
are less than ten percent of the external magnetic field value, so we 
are in the range of validity of the RHMHD model.

Periodic boundary conditions are assumed in all  directions of a cube 
of side $2\pi L$ (where $L\sim 1$ is the initial correlation length of 
the fluctuations, defined as the length unit). The runs performed 
throughout this paper do not contain any magnetic or velocity external 
stirring terms, so the RHMHD system is let to evolve freely.

To generate the initial conditions, we excite initially Fourier modes 
(for both magnetic and velocity field fluctuations) in a shell in $k$-space 
with wavenumbers $1 \le k \le 2$, with the same amplitude for all modes
and with random phases. Only plane-polarized fluctuations (transverse to
the mean magnetic field) are excited, so the initial conditions are
(low- to high-frequency) Alfv\'en mode fluctuations with no 
magnetosonic modes.  

In a first set of simulations, spatial resolution is $512^2$ grid
points in the plane perpendicular to the external magnetic field  
and $32$ grid points in the parallel direction (this is possible 
because the structures that require high resolution only 
develop in the directions perpendicular to the field), allowing four 
different runs to be done with four different Hall coefficients. 
The kinetic and magnetic Reynolds numbers are defined respectively as 
$R=1/\nu$, $R_m=1/\eta$, based on unit initial r.m.s.~velocity 
fluctuation, unit length, and dimensionless values for the viscosity 
and diffusivity. For all the rus, we used $R=R_m=1600$ 
(i.e., $\nu=1/1600$, $\eta=1/1600$). We also considered a Mach 
number $M_S=1/4$, and an Alfv\'en Mach number $M_A=1$.

For $\epsilon$, four values were considered, namely 
$\epsilon=0$ (run A, MHD), $1/32$ (run B), $1/16$ (run C), and $1/8$ 
(run D). As the numerical domain used has size $2 \pi$ (see above),
these values correspond respectively to ion skin depths with
associated wavenumbers $k_\epsilon = \infty$, 32, 16, and 8. Data from 
these simulations is used for the analysis in Sec.~\ref{sect:Analysis}.

To quantify the effect of spatial resolution in the level of 
intermittency, runs A and D were computed also (with the same 
parameters) on a larger grid, with spatial resolution of 
$768^2 \times 32$ grid points. This second set of simulations (namely 
runs A2, with $\epsilon=0$, and D2, with $\epsilon=1/8$) are considered 
later in Sec.~\ref{sect:Resolution}.

\section{\label{sect:Measures}Measures of intermittency}

In order to characterize velocity and magnetic field anisotropy, scaling 
laws and intermittency, we present in the following sections power 
spectra, structure functions, and PDFs of velocity and magnetic field 
increments.

The perpendicular total energy spectrum $E(k_\perp$) is defined as usual, 
summing the power of all (velocity and magnetic) modes in Fourier space 
over cylindrical shells with radius $k_\perp$, with their axis aligned 
with the direction of the guide field.

To compute structure functions and PDFs, field increments must be first 
defined. Given the presence of the external magnetic field, it is 
natural to consider an axisymmetric decomposition for the increments. 
In general, the longitudinal increments of the velocity and magnetic 
fields are defined as:
\begin{equation}
\delta u({\bf x},{\bf l})=\left[{\bf u}({\bf x}+{\bf l})-{\bf u}({\bf x}) 
  \right] \cdot \frac{{\bf l}}{|{\bf l}|}, 
\label{eq:deltav}
\end{equation}
\begin{equation}
\delta b({\bf x},{\bf l})=\left[\bf b ({\bf x}+{\bf l})- \bf b({\bf x}) 
  \right] \cdot \frac{{\bf l}}{|{\bf l}|}, 
\label{eq:deltat}
\end{equation}
where the spatial increment ${\bf l}$ can point in any direction. Structure 
functions of order $p$ are then defined as
\begin{equation}
S^u_{p}({\bf l})=\left\langle \delta u^{p}({\bf x},{\bf l})\right\rangle , 
\label{eq:S}
\end{equation}
for the velocity field, and as
\begin{equation}
S^b_{p}({\bf l})=\left\langle \delta b^{p}({\bf x},{\bf l})\right\rangle ,
\label{eq:Sp}
\end{equation}
for the magnetic field. Here, brackets denote spacial average over all 
values of ${\bf x}$. 

These structure functions depend on the direction of the increment. 
For isotropic and homogeneous turbulence, it is a standard practice to 
average over several directions, to obtain the isotropic component of 
the structure functions (see, e.g., \cite{Arad98,Biferale05,Martin10}). 
Due to the axisymmetry associated with the external magnetic field, in 
our case we will be interested instead only in the increments perpendicular 
to $\hat{z}$. We denote increments in this direction as ${\bf l}_{\perp}$, 
and we follow the procedure explained in \cite{Mininni10,Imazio11} to 
average over several directions of ${\bf l}_{\perp}$.

The method can be described as follows. Velocity and magnetic field 
structure functions were computed from Eqs.~(\ref{eq:S}) and ~(\ref{eq:Sp}) 
using 24 different directions for the increments ${\bf l}$, generated 
by integer multiples of the vectors $(1,0,0)$, $(1,1,0)$, $(2,1,0)$, 
$(3,1,0)$, $(0,1,0)$, $(-1,1,0)$, $(1,2,0)$, $(-2,1,0)$, $(-1,2,0)$, 
$(1,3,0)$, $(-3,1,0)$, and $(-1,3,0)$ (all vectors are in units of grid 
points in the simulations), plus the 12 vectors obtained by multiplying 
them by $-1$. Once these structure functions were calculated, the 
perpendicular structure functions $S^y_p(l_{\perp})$ and $S^b_p(l_{\perp})$ 
were obtained by averaging over these 24 directions in the $xy$ plane.

For all runs, this procedure was applied to $9$ snapshots of the 
velocity and magnetic fields, centered around the time of the peak of 
maximum dissipation (at $t \approx 4.5$), and separated by intervals 
$\Delta t = 0.5$.

For large enough Reynolds number, the structure functions are expected 
to show inertial range scaling, i.e., we expect that for some range of 
scales $S^u_p\sim l_\perp^{\xi_p}$ and $S^b_p\sim l_\perp^{\zeta_p}$, where 
$\xi_{p}$ and $\zeta_{p}$ are respectively the scaling exponents of order 
$p$ of the velocity and magnetic field. As sufficient scale separation 
is needed to determine these exponents, in the following section we show 
scaling exponents for runs A ($\epsilon=0$) and D ($\epsilon=1/8$), as 
these runs have well defined MHD (run A) or HMHD (run D) inertial ranges. 
Runs B and C have the ion skin depth in the middle of the inertial range, 
and each subrange (the MHD subrange and the HMHD subrange) is not 
sufficiently resolved to compute exponents.

The scaling exponents for each snapshot of the fields are obtained 
from the structure functions $S^u_p$ and $S^b_p$ using the least square 
method (extended self-similarity \cite{Benzi93,Benzi93b} is not used to 
estimate the exponents). The values presented in the following section 
correspond to the time average over the $9$ snapshots of each field. As 
the errors in the least square calculation are negligible when compared 
with the variations for each snapshot, the errors in the determination 
of the scaling exponents are estimated by the statistical mean square 
error; e.g., for the magnetic field scaling exponents the error is
\begin{equation}
e_{\zeta_{p}}=\frac{1}{M}\sqrt{\sum_{i=1}^{M}
  \left(\zeta_{p_{i}}-\overline{\zeta_{p}}\right)^{2}},
\label{eq:error}
\end{equation}
where $M=9$ is the number of snapshots of the field used in the analysis, 
$\zeta_{p_{i}}$ is the slope obtained from a least square fit for the 
$i$-th snapshot, and $\overline{\zeta_{p}}$ is the mean value averaged 
over all snapshots.

Finally, to complete the analysis, we consider PDFs of longitudinal 
increments and of derivatives of the perpendicular velocity and magnetic 
fields. In all cases, the PDFs are normalized by their variance, 
and will be shown together with a Gaussian with unit variance as a 
reference.

\section{\label{sect:Analysis}Results}

\subsection{Energy spectrum}

\begin{figure}
\centering
\includegraphics[width=8.3cm]{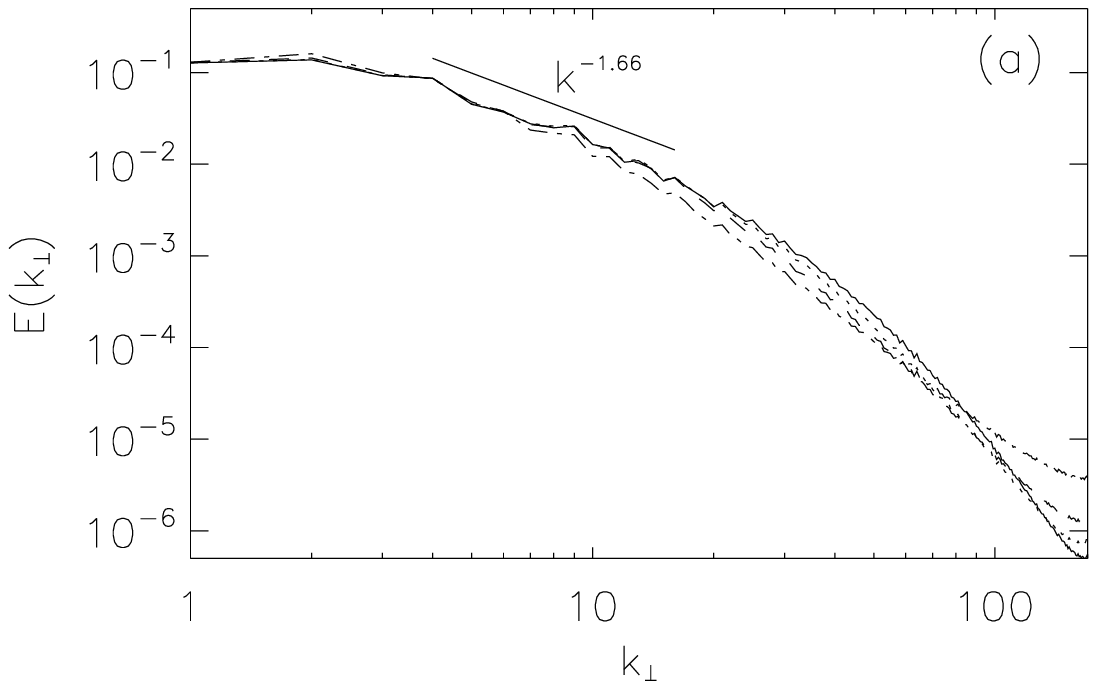}\\
\includegraphics[width=8.3cm]{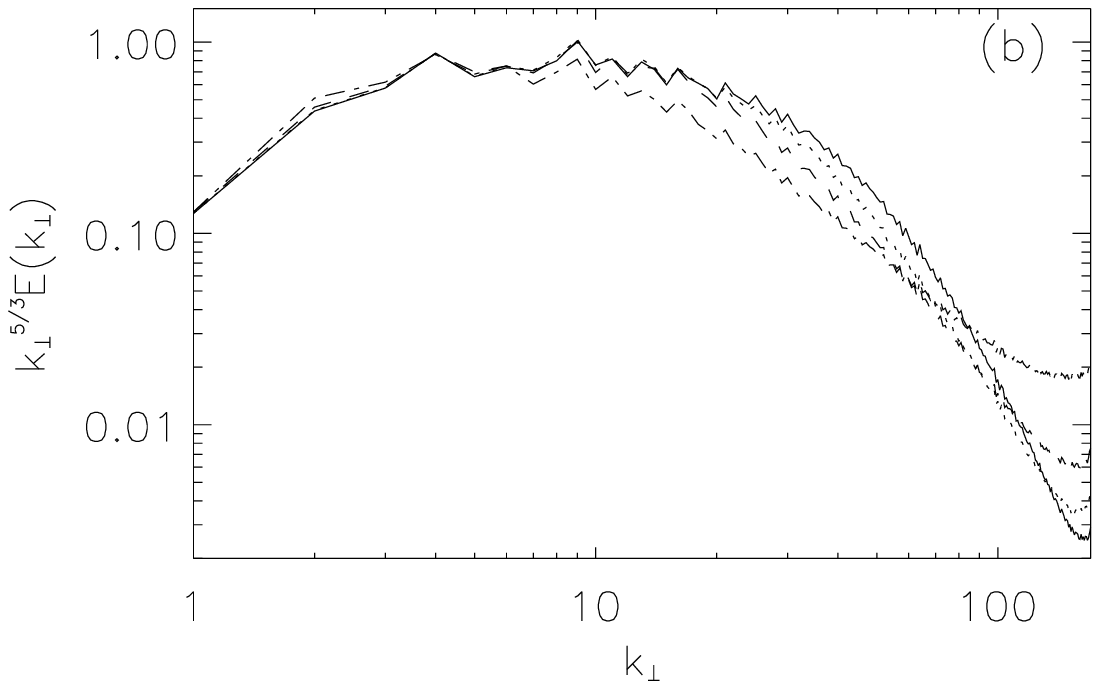}
\caption{(a) Perpendicular energy spectrum for runs A (solid), B 
(dotted), C (dashed), and D (dash-dotted). Note 
how the spectrum becomes steeper in the HMHD simulations for 
wavenumbers larger than the inverse ion skin depth $k_\epsilon$ 
(respectively 32, 16, and 8 for runs B, C, and D). The slope 
indicates Kolmogorov scaling as a reference.
(b) Perpendicular energy spectrum compensated by $k^{-5/3}$ for the 
same runs.}
\label{fig:fig1}
\end{figure}

Before proceeding with the analysis of intermittency, we briefly present 
the total energy spectrum for all the runs with spatial resolution of 
$512^2 \times 32$ grid points. This is important as determination of 
the inertial range based on the spectrum and on the structure functions 
is needed to compute the scaling exponents of the fields.

Figure \ref{fig:fig1} shows the perpendicular spectrum for the 
total energy (kinetic plus magnetic) in runs A, B, C, and D. In run A, 
a range of wavenumbers following an approximate power law can be 
identified, namely from $k\approx 4$ to $k\approx 20$. As a reference, 
we show a Kolmogorov slope. However, it should be noted that 
determination of the slope of the MHD energy spectrum is beyond 
the interest of this work, and readers interested in the topic 
are referred to detailed recent studies on the subject 
\cite{Dmitruk03,Mininni07b,Lee09,Perez12}. 

As the value of $\epsilon$ is increased (see runs B, C, and D in 
Fig.~\ref{fig:fig1}), the spectrum becomes steeper at wavenumbers 
larger than $k_\epsilon$. This has been observed before in simulations 
\cite{Ghosh96,Galtier07,Gomez10}, and it has been argued that it can 
result in an inertial range in the HMHD subrange of the form 
$E(k) \sim k^{-7/3}$ \cite{Galtier08,Alexandrova08}. Run D has a 
HMHD subrange wide enough to compute structure functions and scaling 
exponents, while runs B and C are intermediate between run A and D and 
have two barely resolved subranges. However, these two intermediate 
runs will be useful to study trends in the behavior of the PDFs and 
of the structure functions as $\epsilon$ is increased.

\begin{figure}
\centering
\includegraphics[width=8.3cm]{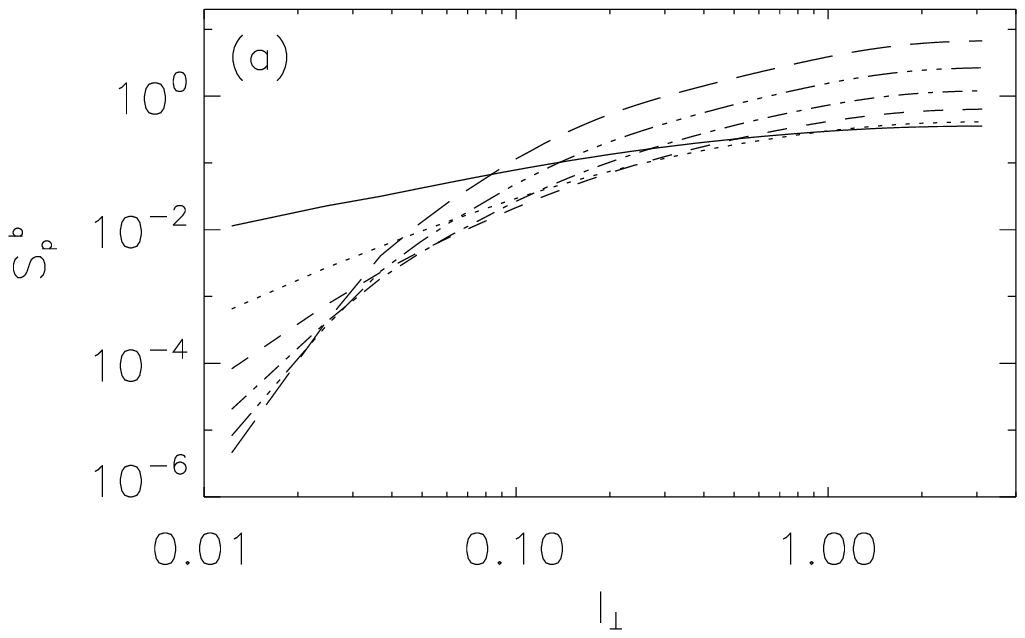}\\
\includegraphics[width=8.3cm]{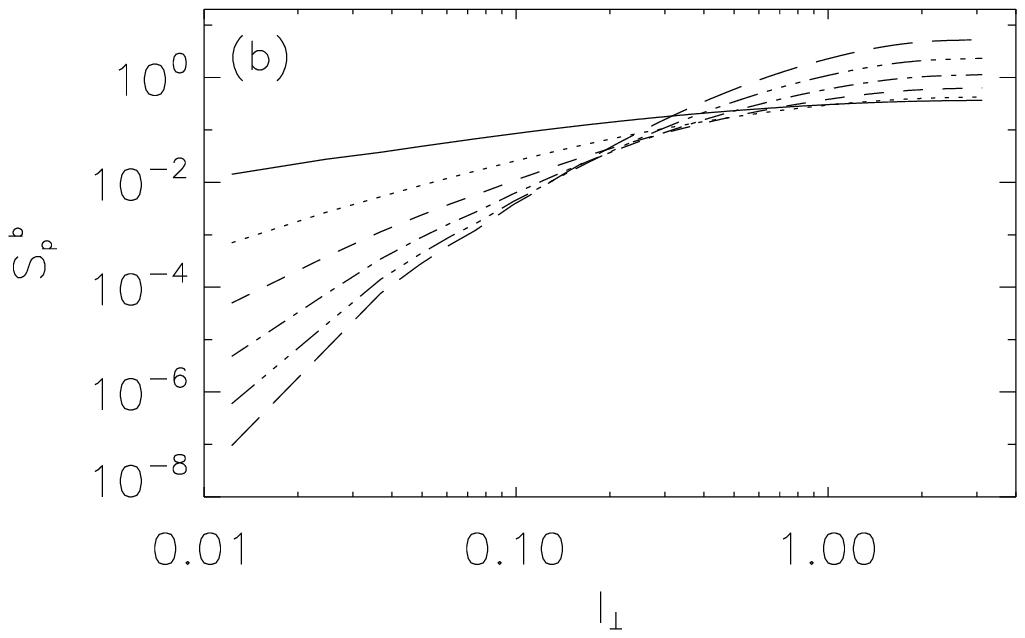}
\caption{Axisymmetric structure functions for the longitudinal 
magnetic field up to six order for (a) run A ($\epsilon=0$), and (b) 
run D ($\epsilon=1/8$). The order of the structure function is 
indicated as follows: $p=1$ (solid), 2 (dotted), 3 (dashed), 4 
(dash-dotted), 5 (dash-triple-dotted), and 6 (long dashes).}
\label{fig:fig2}
\end{figure}

\subsection{Structure functions and scaling exponents}

We present here the results for the computation of the axisymmetric 
structure functions for the longitudinal component of the velocity 
and magnetic field for runs A, B, C, and D.

Figure \ref{fig:fig2} shows the structure functions for the magnetic 
field fluctuations up to sixth order for both runs. The structure 
functions show a range of scales with approximately power law scaling 
at intermediate scales, and at the smallest scales approach the 
$\sim l^p$ scaling expected for a smooth field in the dissipative 
range. The velocity field structure functions (not shown) display a 
similar behavior, at the same range of scales. The inertial range 
identified in the energy spectrum $E(k_\perp)$ is consistent with the 
range of scales where $S^u_p$ and $S^b_p$ show an approximate power law 
behavior.

\begin{figure}
\centering
\includegraphics[width=8.3cm]{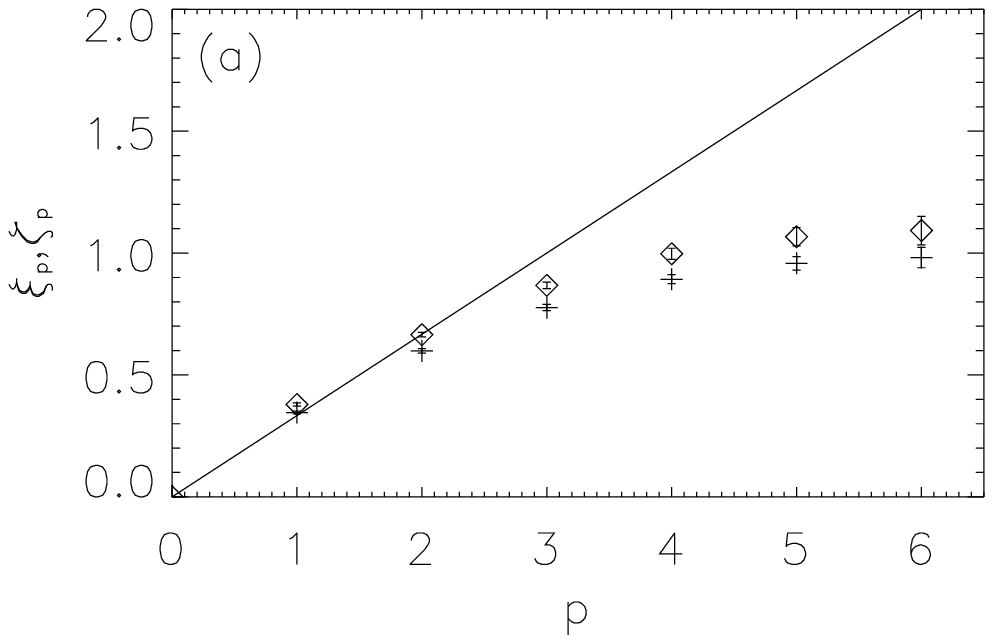}\\
\includegraphics[width=8.3cm]{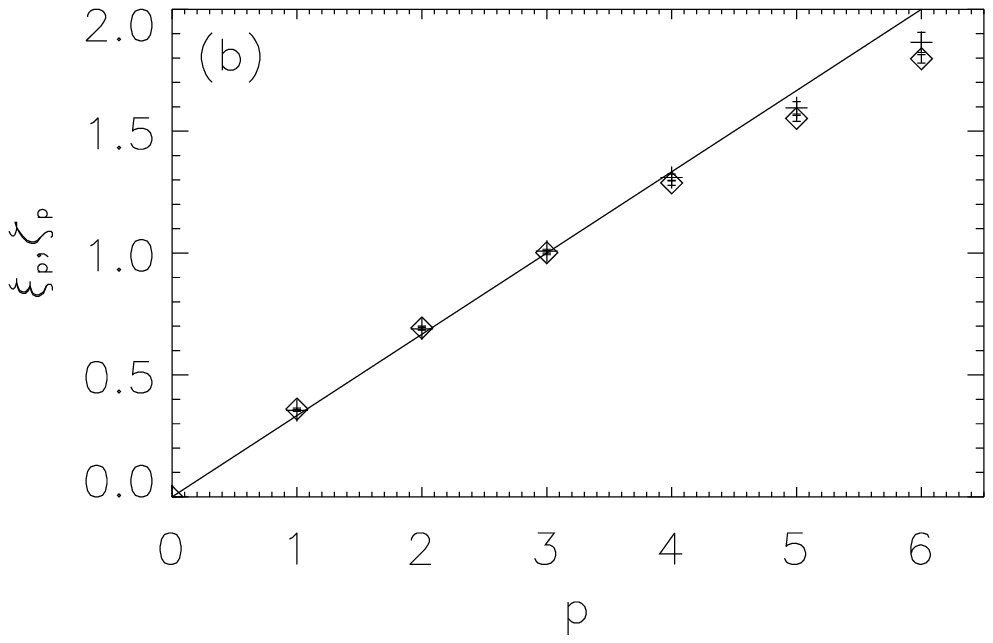}
\caption{Scaling exponents (with error bars) as a function of the 
order $p$ up to sixth order, for the velocity (crosses), and for the 
magnetic field (diamonds), (a) for run A ($\epsilon=0$), and (b) for 
run D ($\epsilon=1/8$). Linear scaling of the exponents with $p/3$ 
(corresponding to non-intermittent scaling with the second order 
exponent consistent with the scaling of the energy spectrum in 
Fig.~\ref{fig:fig1}) is indicated in both cases by the straight line.}
\label{fig:fig3}
\end{figure}

From the structure functions, the scaling exponents can be computed. 
Exponents for the velocity and the magnetic field up to sixth order in 
runs A and D are shown in Fig.~\ref{fig:fig3}. For $\epsilon=0$ (run A) 
the deviation of the exponents $\xi_p$ and $\zeta_p$ from a straight 
line are an indication of intermittency and of multi-fractality. In
the HMHD case ($\epsilon=1/8$, run D), the exponents are closer to 
a straight line, indicating less intermittency. In fact, within error
bars and up to $p=4$, the data is consistent with $\xi_p = \xi_1 p$ and 
$\zeta_p = \zeta_1 p$, and therefore with monoscaling as also observed
for high-frequency magnetic fluctuations in the solar wind 
\cite{Kiyani09}.

The deviation from strict scale invariance (linear scaling) can be 
quantified in terms of the intermittency exponents 
$\mu^u=2\xi_{3}-\xi_{6}$ and $\mu^b=2\zeta_{3}-\zeta_{6}$. The 
larger these exponent, the more intermittent the fields. For run A 
these exponents are $\mu^u=0.57 \pm 0.07$ for the velocity field, 
and $\mu^b=0.64 \pm 0.08$ for the magnetic field. It is interesting 
to point out that these values, that indicate that the magnetic field
is more intermittent than the velocity field, are consistent with
observations of large-scale fluctuations in the solar wind (see, e.g., 
\cite{Podesta07}), and with numerical simulations of MHD turbulence 
at higher spatial resolution \cite{Mininni09}.

The intermittency exponents are substantially reduced
for run D, with $\mu^u=0.15 \pm 0.06$ for the velocity field and 
$\mu^b=0.21 \pm 0.03$ for the magnetic field. This confirms that 
intermittency is substantially decreased in the presence of the Hall 
effect.

\begin{figure}
\centering
\includegraphics[width=8.3cm]{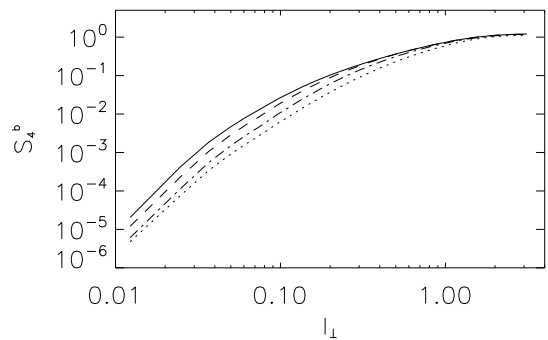}
\caption{Fourth order structure function of longitudinal magnetic 
field increments for runs A ($\epsilon=0$, solid line), B 
($\epsilon=1/32$, dashed), C ($\epsilon=1/16$, dash-dotted), 
and D ($\epsilon=1/8$, dotted).}
\label{fig:fig4}
\end{figure}

At the spatial resolution used in these runs, the lack of sufficient 
scale separation in the MHD and HMHD subranges for intermediate 
values of $\epsilon$ does not allow the calculation of scaling 
exponents for runs B and C. However, the structure functions for 
these runs show a behavior intermediate between runs A and D, and 
consistent with the behavior of the spectrum in Fig.~\ref{fig:fig1}. 
In other words, as the Hall coefficient $\epsilon$ is increased, the 
structure functions steepen at scales smaller than the ion-skin depth. 
As an example of this behavior,  Fig.~\ref{fig:fig4} shows the fourth 
order structure function for the magnetic field for runs A, B, C, and
D. Note that runs B and C show a behavior consistent with the behavior
of run A at large scales (scales larger than the ion-skin depth), and
display a steeper slope (compatible with that found for run D) at
smaller scales.

\begin{figure}
\centering
\includegraphics[width=8.3cm]{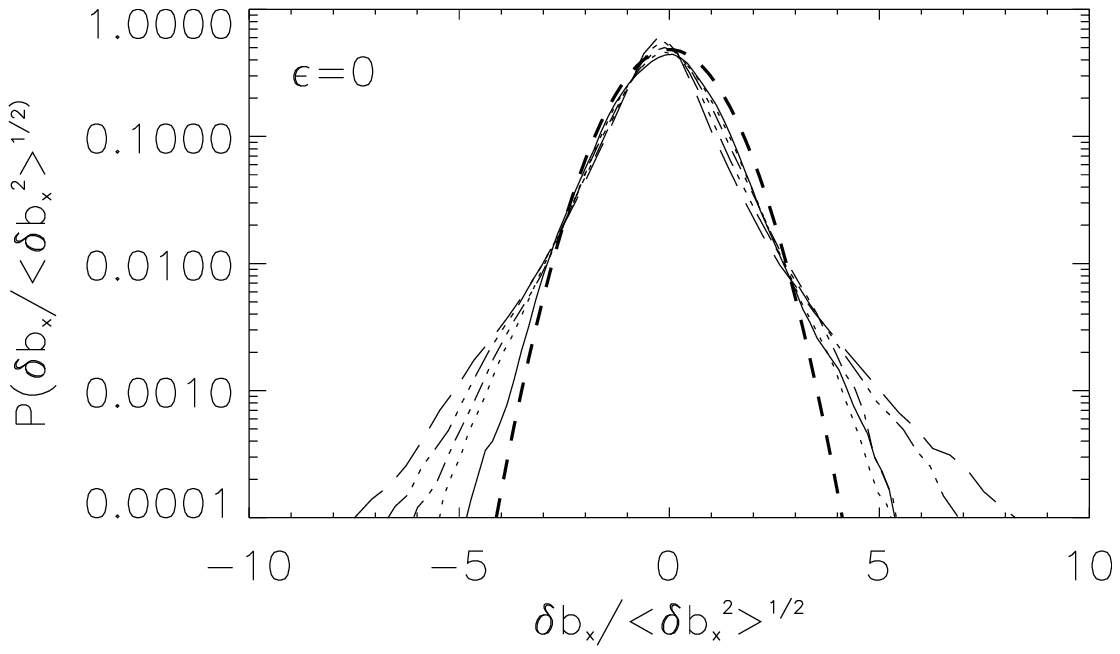}\\
\includegraphics[width=8.3cm]{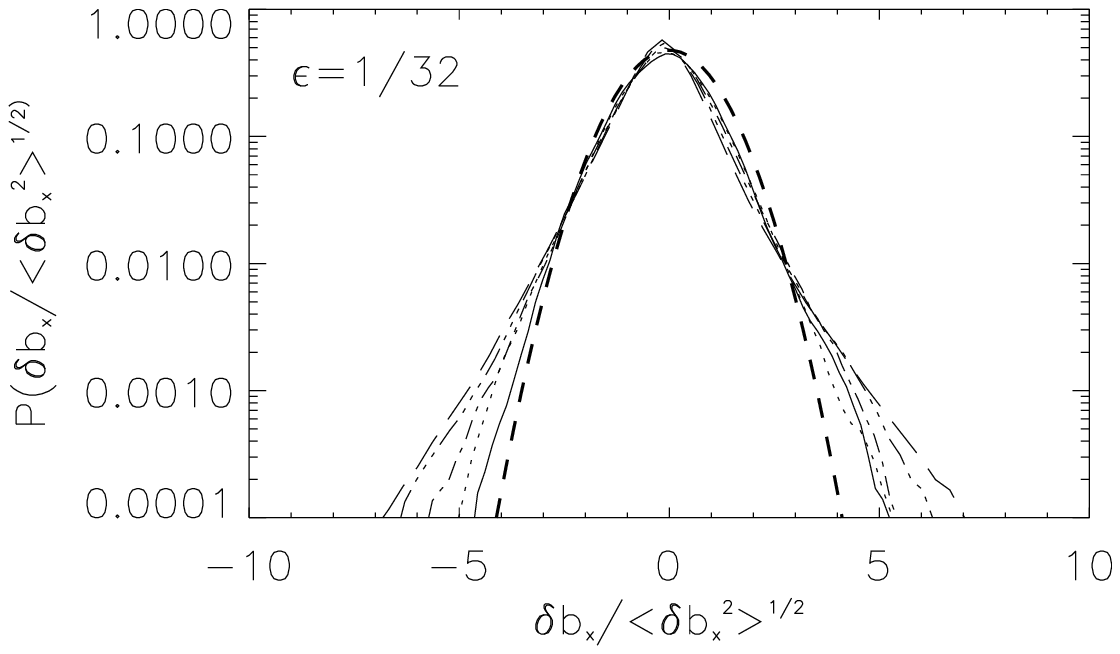}\\
\includegraphics[width=8.3cm]{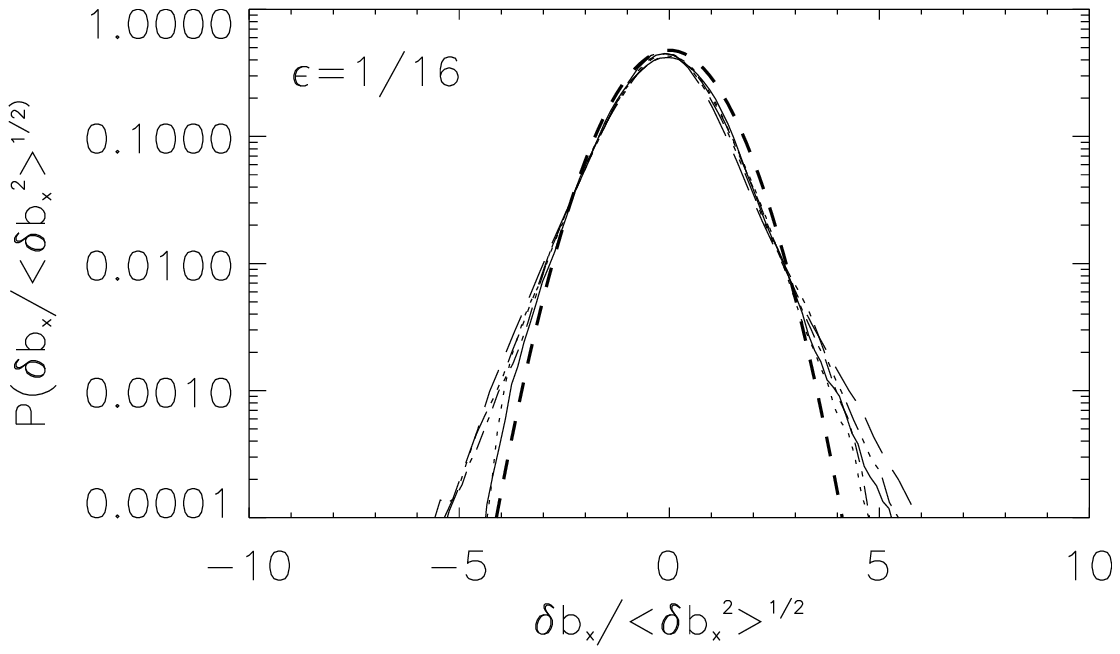}\\
\includegraphics[width=8.3cm]{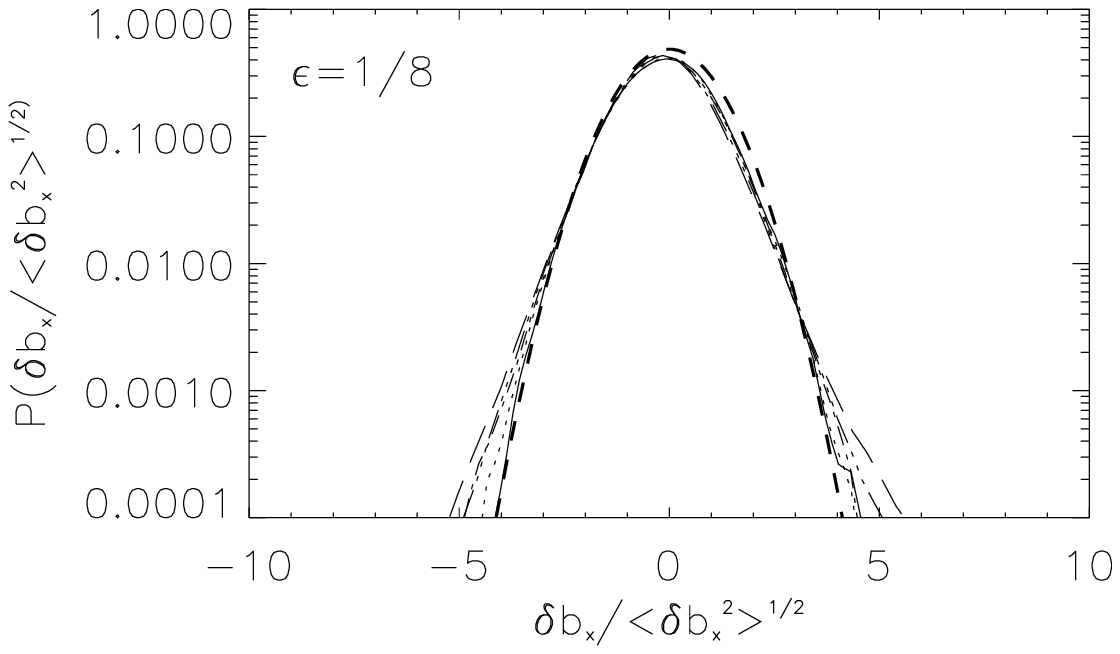}
\caption{PDFs for magnetic field increments, for $l=1.6$ (solid), 
$0.8$ (dotted), $0.4$ (dash-dotted), $0.2$ (dash-triple-dotted), and 
$0.1$ (long dashes), and for runs (a) A, (b) B, (c) C, and (d) D. In
all the figures, a dashed curve indicates a Gaussian PDF with unit
variance.}
\label{fig:fig5}
\end{figure}

\begin{figure}
\centering
\includegraphics[width=8.3cm]{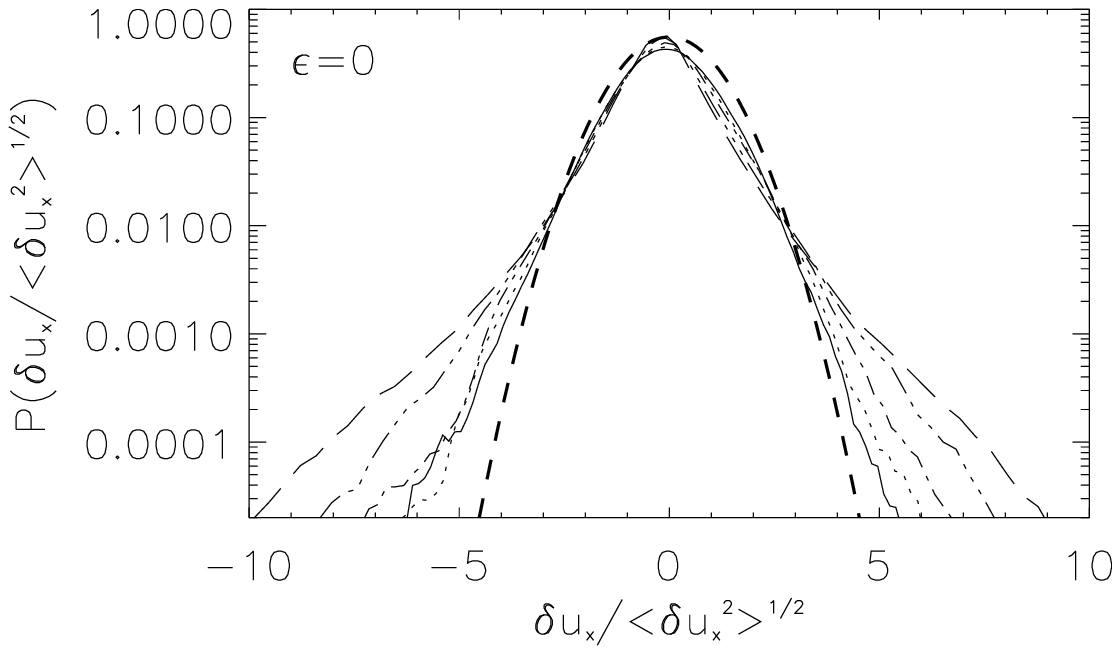}\\
\includegraphics[width=8.3cm]{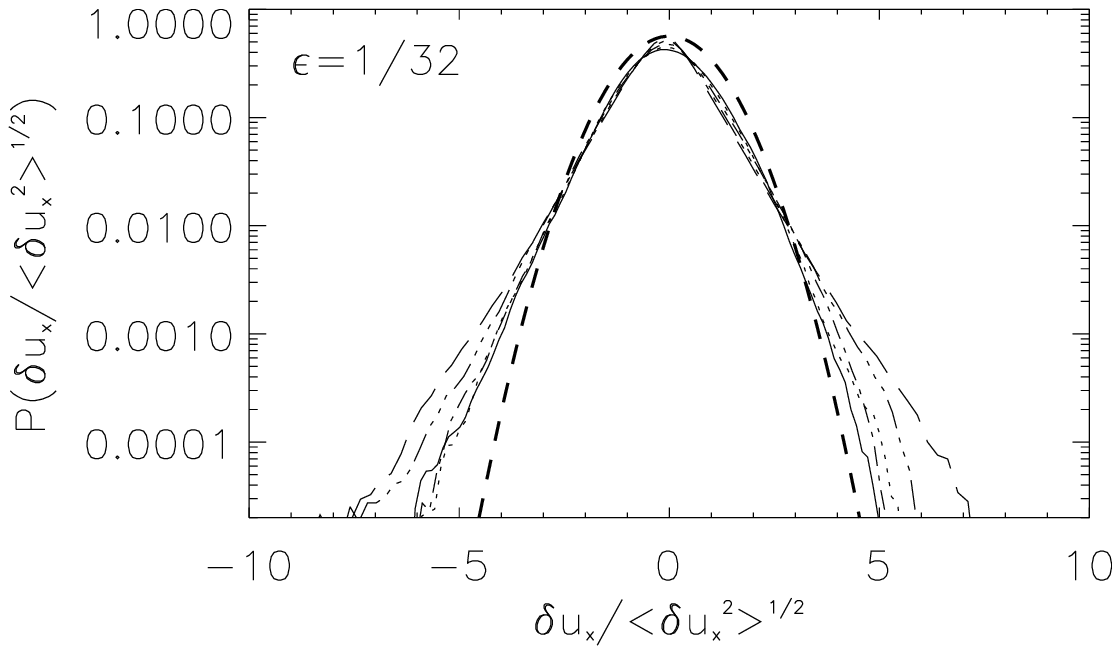}\\
\includegraphics[width=8.3cm]{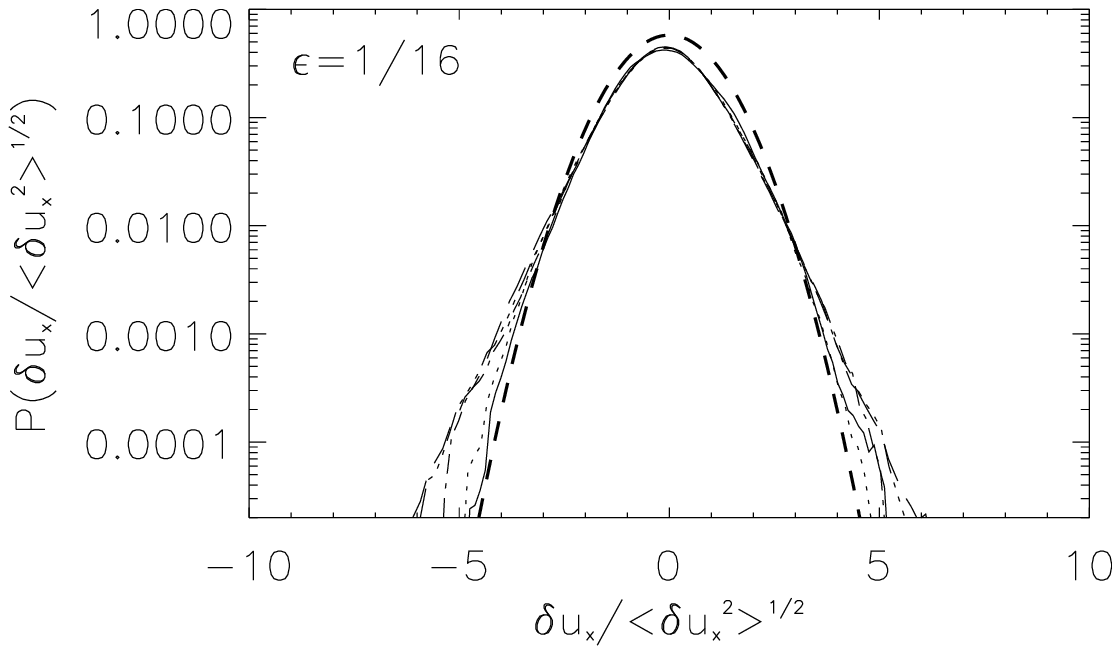}\\
\includegraphics[width=8.3cm]{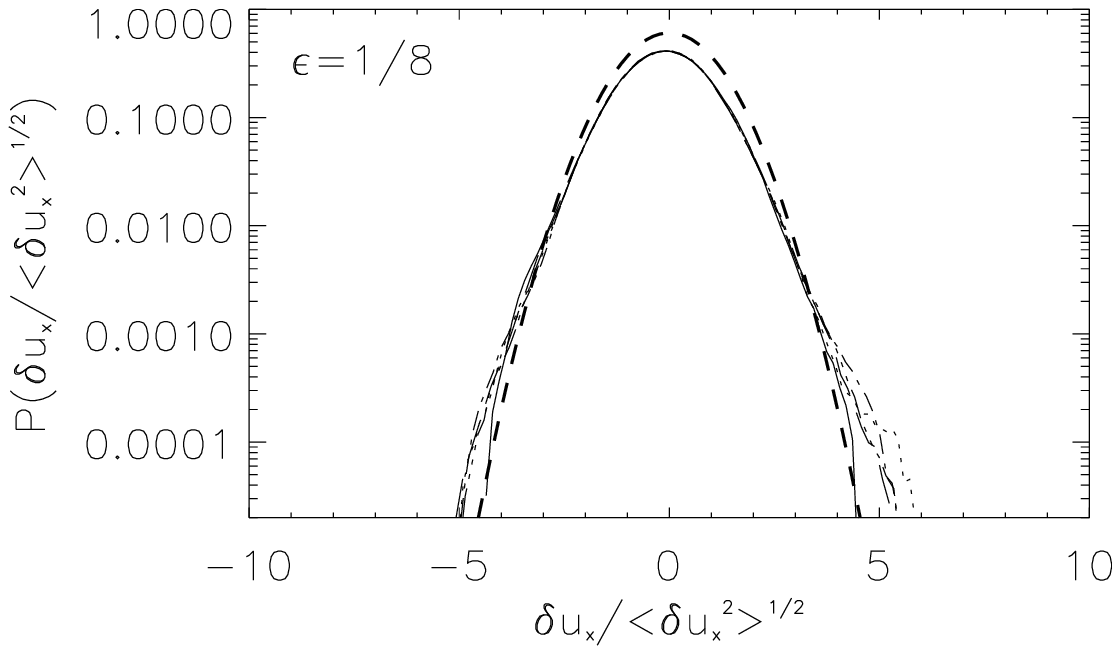}
\caption{PDFs for velocity field increments, for $l=1.6$ (solid), 
$0.8$ (dotted), $0.4$ (dash-dotted), $0.2$ (dash-triple-dotted), and 
$0.1$ (long dashes), and for runs (a) A, (b) B, (c) C, and (d) D. In
all the figures, a dashed curve indicates a Gaussian PDF with unit
variance.}
\label{fig:fig6}
\end{figure}

The results confirm that the presence of the Hall term steepens the 
scaling of the energy spectrum (and consistently, of the structure
functions), and also show that the Hall effect reduces intermittency 
in the velocity and magnetic fields. The velocity and magnetic field 
scaling exponents approach a linear behavior characteristic of a 
self-similar (non-intermittent) flows. In the next section, this
result is confirmed  by an analysis of PDFs of velocity and 
magnetic field increments and spatial derivatives.

\subsection{Probability density functions}

We now consider PDFs for longitudinal increments of the 
$x$-component of the velocity and magnetic fields. As already
mentioned, the PDFs will be presented normalized by their variance, 
and together with a Gaussian distribution with unit variance as a 
reference. Deviations from Gaussianity, or increase of the deviations 
from Gaussianity as smaller increments are considered, are a signature
of intermittency.

\begin{table}
\caption{\label{table:SyK}Skewness ($S$) and kurtosis ($K$) for the 
$x$-derivatives of  $b_x$ and $u_x$, for all runs with spatial
resolution $512^2\times 32$ and with different amplitudes of the Hall
effect $\epsilon$. $S(\partial_{x} b_{x})$ and $K(\partial_{x} b_{x})$
are, respectively, the skewness and kurtosis of the magnetic field 
spatial derivatives, while $S(\partial_{x} u_{x})$ and $K(\partial_{x} u_{x})$ 
are the corresponding quantities for the velocity field derivatives.}
\begin{ruledtabular}
\begin{tabular}{ccccc}
Quantity                  &$\epsilon=0$ &$\epsilon=1/32$ & 
    $\epsilon=1/16$ & $\epsilon=1/8$  \\
\hline
$S(\partial_{x} u_{x})$&$-0.18$&$-0.013$ &$-0.01$ &$-0.001$ \\
$K(\partial_{x}u_{x})$ &$19$     & $8.1$ & $5$ & $4.9$      \\
$S(\partial_{x} b_{x})$&$0.36$  & $0.17$  & $0.11$ & $0.07$ \\
$K(\partial_{x}b_{x})$ &$26$    & $15.7$ & $6.6$ & $5.8$  \\
\end{tabular}
\end{ruledtabular}
\end{table}

Figure \ref{fig:fig5} shows the PDFs of the magnetic field increments 
for four different spatial increments, namely $l=1.6$, $0.8$, $0.4$, 
and $0.1$, for runs A, B, C, and D. For all runs, the PDFs of magnetic 
field increments are close to Gaussian for $l=1.6$,  while for smaller 
spatial increments non-Gaussian tails and asymmetry develop. This is 
a common feature for many turbulent flows, with large scales close to 
Gaussian statistics and smaller scales developing deviations from 
Gaussianity with strong tails (i.e., with extreme gradients more 
probable than what can be expected from a normal distribution). As a 
reference, the integral scale in all runs (the scale with the energy
containing eddies) is close to the size of the domain, $L \approx 2\pi$,
while the dissipative scale is $L_\eta \approx 0.05$. Increments with 
$l=1.6$ are close to the flow integral scale, increments with $l=0.8$ 
or $0.4$ are in the inertial range, while $l=0.1$ is close to the 
dissipation length scale.

Although all runs develop non-Gaussian tails, when comparing the 
PDFs of the four runs with different values of $\epsilon$, it is clear 
that the amplitude of these tails is drastically reduced as the value
of $\epsilon$ is increased. Moreover, for the largest value of 
$\epsilon$ considered, we cannot identify a clear increase in the
amplitude of the tails as we look at smaller increments. This tendency
(which is monotonic with increasing $\epsilon$) of the PDFs of
different spatial increments to collapse into a single curve, with
weaker tails than in the MHD case, is an indication of reduced
intermittency and expected for scale-invariant flows.

Figure \ref{fig:fig6} shows the same PDFs for increments of the
velocity field. Again, the PDFs are close to Gaussian for the largest
increment in the four runs, and non-Gaussian tails develop with 
increasing amplitude for smaller increments. In this case, for 
$\epsilon = 1/8$ all the PDFs seem to collapse into the Gaussian, and
the tails are weaker than for the magnetic field. This is consistent
with the previous observation, using the intermittency coefficients 
$\mu^u $ and $\mu^b$, that the magnetic field is more intermittent
than the velocity field, and that both fields are less intermittent 
in HMHD than in MHD.

\begin{figure}
\centering
\includegraphics[width=8.3cm]{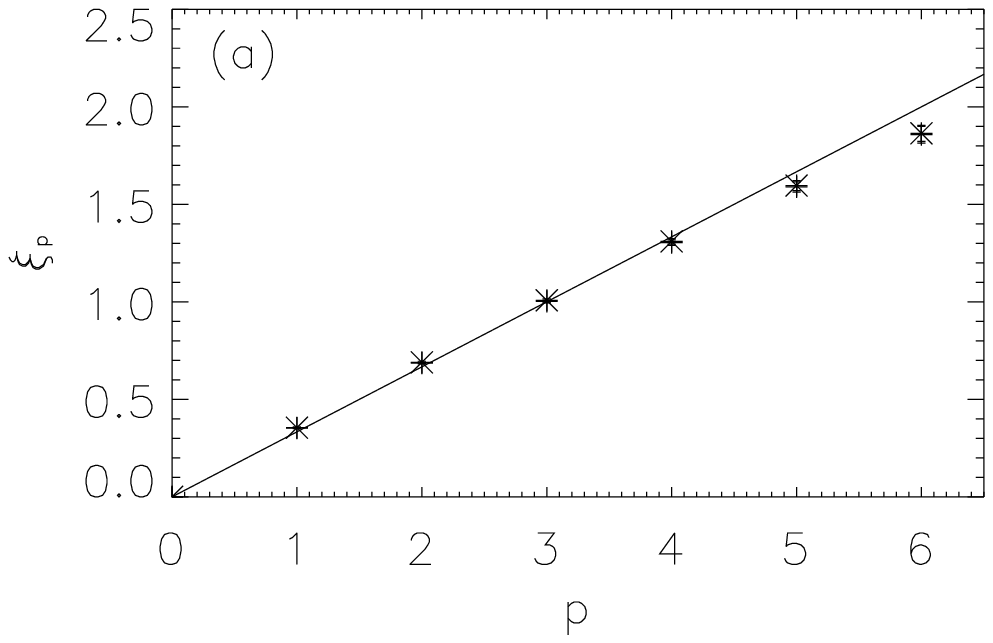}
\includegraphics[width=8.3cm]{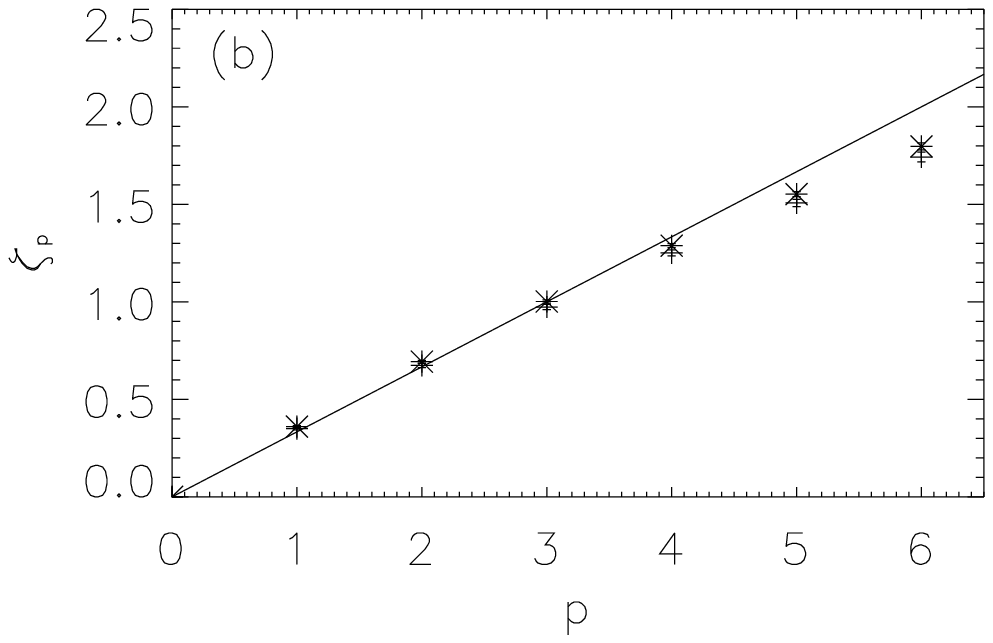}
\caption{(a) Velocity field scaling exponents (with error bars) as a 
function of the order $p$ up to sixth order, for runs D (stars) and D2 
(crosses), both with $\epsilon=1/8$. Linear scaling of the exponents 
is indicated as a reference. (b) Same for the magnetic field scaling 
exponents.}
\label{fig:fig9}
\end{figure}

To quantify the deviations from a Gaussian distribution in each run, 
we calculated the skewness and the kurtosis of the $x$-derivatives 
of the $x$-components of the velocity and magnetic fields. Note these 
quantities correspond respectively to the third- and fourth-moments of 
the PDFs in Figs.~\ref{fig:fig5} and ~\ref{fig:fig6} in the limit of vanishing 
spatial increment. The skewness and kurtosis of a function $f$ are 
defined as 
$S(f)=\left< f^3 \right> / \left< f^2\right>^{3/2}$ 
and
$K(f)=\left< f^4 \right> / \left< f^2 \right>^{2}$ respectively, where 
$f$ can be, e.g., some component of the velocity (or magnetic) field 
gradient. The resulting values are listed in Table \ref{table:SyK}. In
accordance with what can be expected from a visual inspection of 
Figs.~\ref{fig:fig5} and ~\ref{fig:fig6}, the skewness of 
$\partial _{x}u_{x}$ and $\partial _{x}b_{x}$ is reduced to almost 
zero for $\epsilon=1/8$, which indicates a substantial reduction in
the asymmetry of the PDF. The kurtosis of $\partial _{x}u_{x}$ and 
$\partial _{x}b_{x}$ also decreases with increasing $\epsilon$, wich 
indicates a smoothing in the peakedness of the PDFs and a decrease 
in the intensity of the tails.

\begin{figure}
\centering
\includegraphics[width=8.3cm]{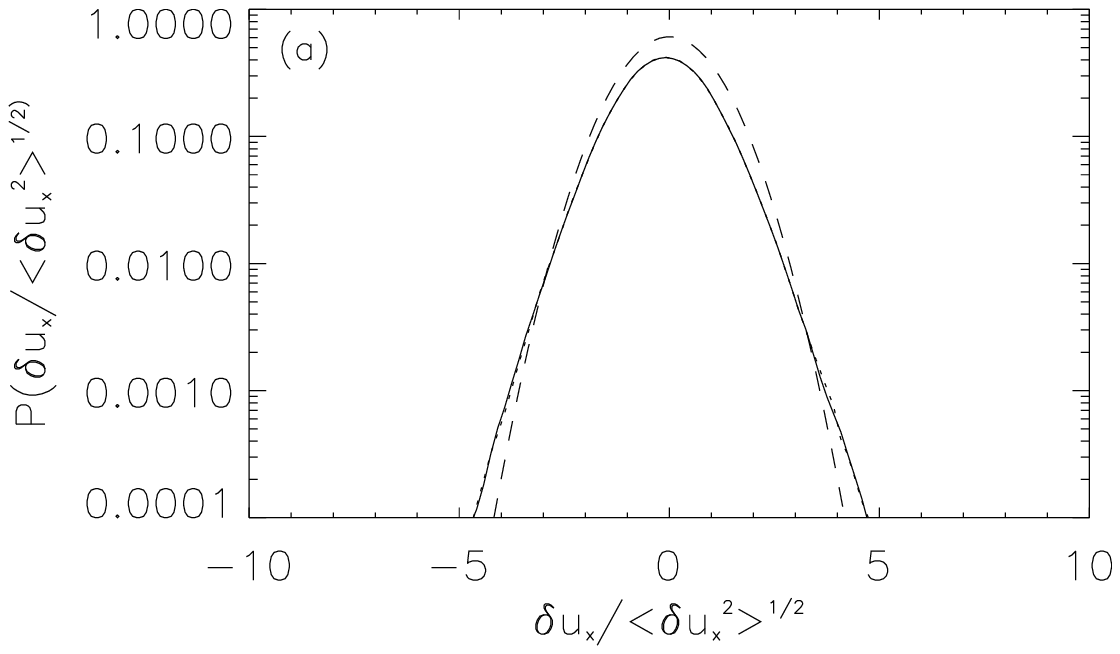}
\includegraphics[width=8.3cm]{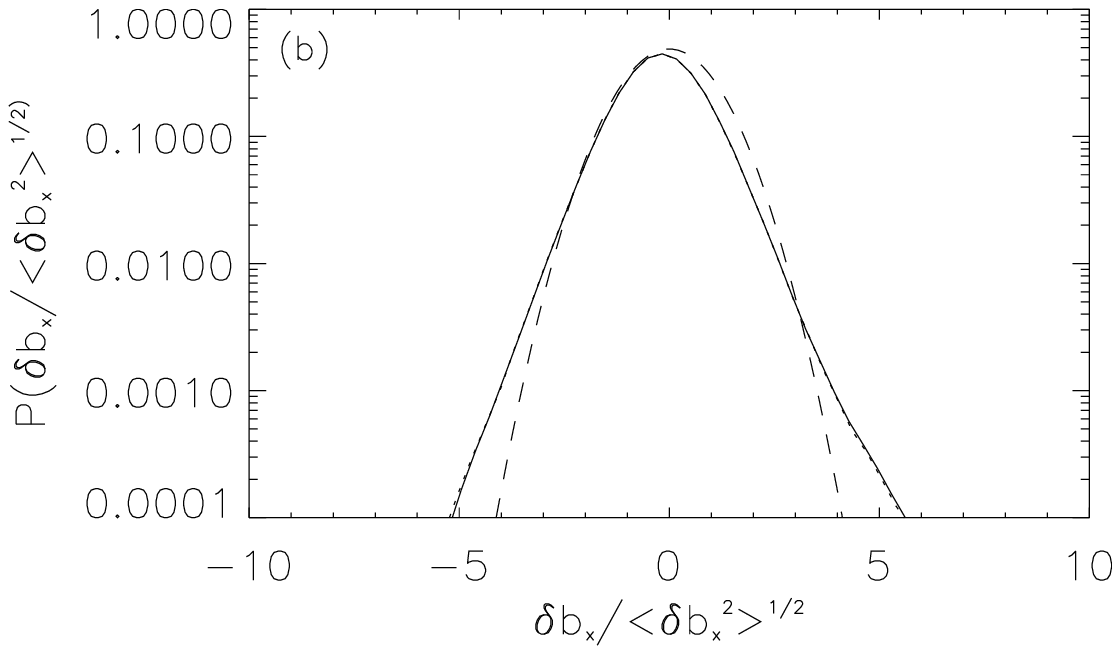}
\caption{(a) PDFs of velocity field increments for $l=0.1$ and 
$\epsilon=1/8$, for runs D (solid line) and D2 (dotted). 
The two PDFs are practically indistinguishable. The dashed 
line shows a Gaussian distribution as a reference. (b) Same for 
magnetic field increments.}
\label{fig:fig13}
\end{figure}

\section{\label{sect:Resolution}Effect of resolution}

Recently, it was stressed the need of using well resolved 
numerical simulations to quantify high order statistics and 
intermittency in MHD \cite{Wan10}. In particular, it has been claimed that 
if the flow is not properly resolved, a partial thermalization of the 
small scales may result in artificial Gaussian statistics and an 
artificial decrease of the intermittency. Considering this, in this 
section we present results for simulations with the same parameters 
as in runs A and D, but with larger spatial resolution 
($768^2 \times 32$ grid points). We will refer to these two runs as 
runs A2 and D2.

We computed structure functions, scaling exponents, and PDFs for runs
A2 and D2 and compared the results with those found for runs A and
D. In all cases, the results were consistent within error bars. As an
illustration, in Fig.~\ref{fig:fig9} we show the velocity field and
magnetic field scaling exponents for runs D and D2 (both with 
$\epsilon=1/8$, the former with $512^2\times 32$ grid points, and 
the later with $768^2\times 32$ grid points). Increasing the
resolution does not change the scaling exponents, nor does it change 
the fact that the exponents are close to the straight line and less 
intermittent than in the MHD case.

In run A2, the intermittency exponents are $\mu^u=0.52 \pm 0.08$ 
and  $\mu^b=0.70 \pm 0.07$, consistent within error bars with the
values found in run A, while in run D2 the intermittency exponents are
$\mu^u=0.15 \pm  0.06$ and $\mu^b=0.20 \pm 0.05$, also 
consistent with the values obtained in run D.

Figure \ref{fig:fig13} shows the PDFs of velocity and magnetic field 
increments in runs D and D2, for a spatial increment $l=0.1$. The 
PDFs are almost indistinguishable. Similar results were obtained for 
runs A and A2. When computing the PDFs of spatial derivatives of 
the fields, we obtained $S(\partial_{x} u_{x})=-0.19$, 
$S(\partial_{x} b_{x})=0.41$, $K(\partial_{x}u_{x})= 18$, and 
$K(\partial_{x}b_{x})= 26$ for run A2, and $S(\partial_{x} u_{x})=-0.01$, 
$S(\partial_{x} b_{x})=0.08$, $K(\partial_{x}u_{x})= 8.4$, and 
$K(\partial_{x}b_{x})= 6.5 $ for run D2 (compare with the values in 
Table \ref{table:SyK} for the runs at lower resolution).

Wan et al. \cite{Wan10} argue that for an MHD simulation to be well 
resolved, the kurtosis of the current should remain independent of the 
spatial resolution. In our MHD and HMHD runs that condition is 
fulfilled, at least up to the level of statistical fluctuations that
can be expected when comparing two simulations of a turbulent flow. 
To verify this, we computed the skewness and kurtosis of the 
component of the current density parallel to the external magnetic 
field, i.e., $S(j_z)$ and $K(j_z)$. In the MHD simulations
($\epsilon=0$), we obtained $S(j_z)=0,70$ and $K(j_z)=21$ in the 
simulation with $512^2 \times 32$ grid points, and $S(j_z)=0,71$ 
and $K(j_z)=22$ in the simulation with $768^2 \times 32$ grid 
points. In the HMHD simulations with $\epsilon=1/8$, we obtained 
$S(j_z)=-0,02$ and $K(j_z)=4.5$ in the simulation with 
$512^2 \times 32$ grid points, and $S(j_z)=-0,01$ and $K(j_z)=4.8$ 
in the simulation with $768^2 \times 32$ grid points.

Although there is a small increase in $S(j_z)$ and $K(j_z)$ as the 
resolution is increased in both the MHD and HMHD runs, the increase 
is smaller than $10\%$ in most cases. As a result, we conclude that 
the simulations are well resolved even with the more stringent
criteria of Wan et al. \cite{Wan10}. Moreover, the reduction of the 
intermittency in presence of the Hall term is also confirmed by the 
skewness and kurtosis of the current at both spatial resolutions.

As a result, we conclude that increasing resolution has no significant
effect on the results we reported in the previous section, and that the
decrease in the intermittency of the flow presented above has its
source in the Hall effect and not in a numerical artifact when the
flow is not properly resolved.

\section{Summary and conclusions}

In this work, we presented a study of intermittency in the velocity
and magnetic field fluctuations of compressible 
Hall-magnetohydrodynamic turbulence with an external guide field. 
Unlike previous works, we were not interested in the characterization 
of geometrical properties or in the size of individual structures in the 
flow (e.g., current sheets), but rather interested in their overall statistical 
properties. 

The equations were solved numerically using a reduced model valid 
when a strong guide field is present, and both structure functions and 
probability density functions of field increments were computed. In 
the magnetohydrodynamic limit we recovered results found in previous 
studies, with the magnetic field being more intermittent than the
velocity field. However, in the presence of the Hall effect, we found
field fluctuations at scales smaller than the ion skin depth to be
substantially less intermittent, with close to scale-invariant scaling.

As the intensity of the Hall effect was increased in the simulations
(i.e., the ion skin depth was made larger in units of the box size),
we found both the total energy spectrum and the structure functions to 
develop a steeper scaling in a wider subinertial range, for all scales
smaller than the ion skin depth. The behavior of the scaling exponents 
for both the velocity and the magnetic field up to sixth order becomes 
closer to monofractal as the Hall effect is increased, and the
intermittency exponent decreases accordingly.

In agreement with these results, the probability density functions of 
longitudinal velocity and magnetic field increments have weaker
non-Gaussian tails and less asymmetry at scales smaller than the ion
skin depth. For velocity and magnetic field gradients, the skewness
and kurtosis also decreases as the Hall effect is increased.

These results were obtained for simulations with spatial resolution of 
$512^2 \times 32$ grid points, and verified in simulations at larger 
spatial resolution, with $768^2 \times 32$ grid points. As a result, 
we can safely conclude that increasing resolution has no effect on
the results, and that the decrease in the intermittency of the flow
has its source in the Hall effect.

\begin{acknowledgments}
PD and PDM acknowledge support from the Carrera del Investigador 
Cient\'{\i}fico of CONICET. The authors acknowledge support from 
grants PICT 2011-1626 and 2011-1529, PIP 11220090100825, and 
UBACYT 20020110200359.
\end{acknowledgments}

\end{document}